\documentclass[12pt]{article}

\usepackage{amsfonts, graphicx,caption,subfig}

\usepackage{amssymb}
\usepackage{amsmath}
\usepackage{amsfonts}

\def\be{\begin{equation}}
\def\ee{\end{equation}}
\def\ben{\begin{displaymath}}
\def\een{\end{displaymath}}
\def\ba{\begin{array}{c}}
\def\ea{\end{array}}
\def\p{\partial}

\newcommand{\bea}{\begin{eqnarray}}
\newcommand{\eea}{\end{eqnarray}}

\newcommand{\kt}{\rangle}
\newcommand{\br}{\langle}
\newcommand{\ed}{\end{document}}
\newcommand{\bbr}{\br\!\br}
\newcommand{\kkt}{\kt\!\kt}

\begin{document}

\begin{center}

{\Large \bf

Crypto-unitary forms of quantum evolution operators

  }

\vspace{9mm}

{Miloslav Znojil}

\vspace{9mm}

Nuclear Physics Institute ASCR, 250 68 \v{R}e\v{z}, Czech Republic

{znojil@ujf.cas.cz}

{http://gemma.ujf.cas.cz/\~{}znojil/}

\end{center}

\section*{Abstract}

The description of quantum evolution using unitary operator
$\mathfrak{u}(t)=\exp(-{\rm i}\mathfrak{h}t)$ requires that the
underlying self-adjoint quantum Hamiltonian $\mathfrak{h}$ remains
time-independent. In a way extending the so called  ${\cal
PT}-$symmetric quantum mechanics to the models with manifestly
time-dependent ``charge'' ${\cal C}(t)$ we propose and describe an
extension of such an exponential-operator approach to evolution to
the manifestly time-dependent self-adjoint quantum Hamiltonians
$\mathfrak{h}(t)$.

\newpage

\section{Introduction \label{oh} }

It is well known from textbooks \cite{Messiah} that quantum theory
describes the unitary evolution of a system in time via its
self-adjoint generator $\mathfrak{h}=\mathfrak{h}^\dagger$ called
Hamiltonian. Thus, in principle, one prepares a state (i.e., an
element $|\varphi(t)\kt$ of a physical Hilbert space ${\cal
H}^{(P)}$) at time $t=t_{initial}=0$. Subsequently, one performs a
measurement over the system at a positive $t=t_{final}>0$. Inside
the interval, the time-evolution of the state may be reconstructed
via Schr\"{o}dinger equation
 \be
 {\rm i}\p_t|\varphi(t)\kt = \mathfrak{h}\,|\varphi(t)\kt\,,
 \ \ \ \ \ \
 |\varphi(t)\kt \in {\cal H}^{(P)}\,.
 \label{SEtaune}
 \ee
In practice, our attention remains often restricted to the case of
the stationary models based on the time-independent
$\mathfrak{h}=\mathfrak{h}(0)$ for which the states are described by
the well known operator-exponential formula
 \be
 |\varphi(t)\kt=\exp (-{\rm i}\mathfrak{h}(0)
 \,t)\,|\varphi(0)\kt\,.
 \ee
In such a setting, nontrivial difficulties may only emerge when the
Hamiltonian $\mathfrak{h}$ (which is not allowed to vary with time)
proves prohibitively difficult by itself. For an illustrative
example one may recall the review paper \cite{Geyer} where where
several phenomenological, highly instructive illustrations of such a
scenario have been analyzed in the context of nuclear physics. A few
years later, the similar problem of the practical intractability of
an overcomplicated realistic Hamiltonian $\mathfrak{h}$ re-emerged
in the context of field theory and has been solved in similar manner
(cf., e.g., the review papers \cite{Carl} or \cite{ali}).

The essence of the underlying common theoretical idea of the
potential simplification of the overcomplicated but still
time-independent and self-adjoint Hamiltonian $\mathfrak{h}$ will be
briefly summarized in section \ref{ho} below. Its core will be shown
to lie in the replacement of $\mathfrak{h}$ by its
isospectral-partner representation
 \be
 H=\Omega^{-1} \mathfrak{h}\Omega\,.
 \label{ofeq}
 \ee
Naturally, once one follows such a recipe and tries to replace a
complicated operator  $\mathfrak{h}$ by its sufficiently simplified
alternative $H$, a number of problems emerges in connection with the
search for the suitable mapping $\Omega$.

In the literature one finds, in essence, two alternative strategies
of avoiding such a trap. Firstly, in a way exemplified in
\cite{Geyer} one starts form the knowledge of a prohibitively
complicated but still well-defined (i.e., typically, realistic and
microscopic) Hamiltonian $\mathfrak{h}=\mathfrak{h}^\dagger$. By the
method of trial and error (based, usually, on some additional,
physics-based knowledge about the system in question) one then tries
to select a suitable operator $\Omega$. In the third step of the
algorithm one finally checks the required gain in simplicity,
typically, by checking the amendment of the rate of the practical
numerical convergence of the eigenvalues $E_n$ when calculated from
$H$ \cite{Geyer}.

The second methodical alternative is well known under the nickname
of  ${\cal PT}-$symmetric quantum mechanics \cite{Carl} or, in an
inessentially more universal formulation, of the pseudo-Hermitian
quantum mechanics \cite{ali}. In this approach one starts from a
suitable and, by assumption, sufficiently elementary second
representation $H$ of the realistic Hamiltonian. Subsequently one
reconstructs the bound-state spectrum $E_n$ and compares it with the
experimental or phenomenological data (if any) immediately.

In the latter (let us conventionally call it, for the time being,
${\cal PT}-$symmetric) approach, the reconstruction of the original,
``true'' Hamiltonian $\mathfrak{h}$ is often being postponed to the
very end of all of the considerations. This has, in principle, two
rather unpleasant consequences. Firstly, one usually encounters
rather serious technical \cite{otherpap} as well as conceptual
\cite{Jones} difficulties with the very physical interpretation of
the ${\cal PT}-$symmetric models. Secondly,  the construction of the
original self-adjoint version $\mathfrak{h}$ of the Hamiltonian
itself becomes almost redundant. In the majority of cases, moreover,
this construction remains just approximative and, in addition, also
remarkably difficult as a rule \cite{cubic}.

The former approach starts form the initial knowledge of
$\mathfrak{h}=\mathfrak{h}^\dagger$ and its main merit is that the
related physical interpretation of the system is without problems.
The success of such an approach (let us conventionally call it, for
the reasons which will be clarified later, crypto-unitary) is then
measured but the success of the trial and error selection of
$\Omega$ leading to a sufficiently friendly (and, in principle,
potentially also  ${\cal PT}-$symmetric or, in the language of
mathematics, Krein-space self-adjoint \cite{Langer}) effective
Hamiltonian $H$ of Eq.~(\ref{ofeq}).

In comparison, the respective merits of these two approaches may be
perceived as complementary  and application-dependent. In parallel,
one of their ``shared'' weaknesses can be seen, in a way explained
and summarized in our paper I \cite{timedep}, in the unnecessary and
mathematically rather artificial above-mentioned requirement of the
time-independence of the individual operators $\mathfrak{h}$, $H$
and/or $\Omega$. In this sense, our present paper may be perceived
as an immediate continuation of paper I \cite{timedep} (cf. also its
conference-proceeding extension \cite{SIGMA}) where we generalized,
consequently, the methods of Refs.~\cite{Geyer,Carl,ali} to the
class of quantum models where the initial (i.e., presumably,
complicated) Hamiltonian of Eq.~(\ref{SEtaune}) becomes allowed {\em
manifestly time-dependent}, i.e., where
$\mathfrak{h}=\mathfrak{h}(t)$ for $t \in (t_{initial},t_{final})$.

The readers may perceive our present paper as motivated by the
difficulties encountered during the attempted solution of
Schr\"{o}dinger Eq.~(\ref{SEtaune}) with the Hamiltonian
$\mathfrak{h}=\mathfrak{h}(t)$. In other words, we shall offer here
a continuation of paper I in which we shall develop further the very
pragmatic attitude of Ref.~\cite{Geyer} where the key purpose of the
whole approach has been emphasized to lie in the perceivable {\em
simplification} of practical calculations. In this sense we may
formulate our present aim as the statement of a {\em possibility of
an elimination} of the manifest time-dependence from the properly
simplified version of the generator of quantum evolution.

The mathematical motivation of such a project may be traced back to
the unexpected emergence of a few rather serious obstacles which
have been encountered during attempted implementations of the
generalized formalism of paper I. {\it Pars pro toto}, we found it
rather unpleasant that virtually all of these applications appeared
to require an additional simplification of technicalities mediated,
typically, by the choice of a trivial time-dependence in
$\mathfrak{h}(t)$ \cite{Fring,alitimedep} or by the use of various
versions of adiabatic-approximation hypothesis \cite{Bila,Bang}.

The key technical ingredients of our present proposal will make use
of the details explained in paper I. We shall recall also
Ref.~\cite{SIGMA} and, in its spirit, we shall also make use of the
notation of this reference. The presentation of our message will be
separated into a concise review of the existing time-independent
theory (section \ref{ho}) and of its time-dependent completion as
given in paper I (section \ref{hoho}), followed by the description
of the main result (section \ref{hohoho}), by the discussion
(section \ref{hohohoho}) and by a brief summary (section
\ref{hohohohoho}).

\section{Time-independent non-Hermitian quantum Hamiltonians \label{ho} }

The current popularity of non-Hermitian Hamiltonians $H \neq
H^\dagger$ \cite{Hook} grew from multifaceted physical origins
ranging from relativistic quantum field theory \cite{BM} and from
cosmology \cite{Kamenshchik,ali} to nuclear physics
\cite{Geyer,Rotter}, optics \cite{optics}, magnetohydrodynamics
\cite{MHD}, thermodynamics \cite{Vit}, scattering theory
\cite{Jones}, electromagnetism \cite{loran} and  quantum chemistry
\cite{Nimrod}. The mathematical and formal aspects of these
innovative applications involve, in the context of the very
traditional quantum theory, perturbation analysis \cite{Caliceti},
analytic continuations \cite{BG}, the calculus of variations
\cite{Geyer}, supersymmetry \cite{ptsusy} and the Feshbach's
model-space techniques \cite{Rotter,Nimrod,brachys}.

The profit provided by these developments is a {\em simplification}
of constructive analyses. This inspired an unexpected and powerful
innovation of the traditional model-building strategies. One of the
oldest illustrations of the recipe has been offered via the so
called ``interacting boson models'' \cite{Geyer} where the use of
non-Hermitian phenomenological Hamiltonians $H\neq H^\dagger$
shortened the computer-assisted numerical predictions of the
energy-level spectra of heavy nuclei. Similarly, several
field-theory models appeared tractable {\em solely} in specific
non-Hermitian (a.k.a. ${\cal PT}-$symmetric) versions with $H\neq
H^\dagger={\cal P}H{\cal P}$ where ${\cal P}$ denotes parity
\cite{Carl,cubic}. Last but not least, analogous models found their
innovative applications in cosmology \cite{Kamenshchik}.

At the very beginning of model-building considerations we usually
{\em assume and test} \cite{BBF} (or prove \cite{DDT}) that the
spectrum of $H\neq H^\dagger$ is real and discrete and bounded
below. Under these assumptions we may introduce a family of
isospectral images of the Hamiltonian,
 \be
 \mathfrak{h} = \Omega\,H\,\Omega^{-1}\,.
 \label{simplifya}
 \ee
We may identify them with the Hamiltonians of section \ref{oh} and
require that they are self-adjoint (i.e., observable),
$\mathfrak{h}=\mathfrak{h}^\dagger$. Formally, this merely imposes a
constraint upon the eligible (sometimes called Dyson's \cite{Geyer})
operators $\Omega$,
 \be
 H^\dagger\Theta=\Theta\,H\,,
 \ \ \ \ \ \ \Theta=\Omega^\dagger\Omega\,.
 \ee
In the light of Ref.~\cite{Dieudonne} we may call such a constraint
``Dieudonn\'e's equation''. It may be perceived as a hidden
Hermiticity property or crypto-Hermiticity condition~\cite{SIGMA}.

The latter conclusion makes the core of the whole methodical message
more or less trivial. One merely replaces the standard textbook
Schr\"{o}dinger Eq.~(\ref{SEtaune})  by its, by assumption,
``friendlier'' crypto-Hermitian re-arrangement
 \be
 {\rm i}\p_t|\Phi(t)\kt = H\,|\Phi(t)\kt\,,
 \ \ \ \ \ \
 |\Phi(t)\kt =\Omega^{-1}\,|\varphi(t)\kt \in {\cal H}^{(F)}\,
 \label{SEtautrane}
 \ee
yielding the elementary evolution operator whenever $H\neq H(t)$.
The time-evolution of the friendlier solutions $|\Phi(t)\kt=\exp
(-{\rm i}H \,t)\,|\Phi(0)\kt$ appears non-unitary (unless
$H=H^\dagger$ of course),
 \be
 \br \Phi_1(t)|\Phi_2(t)\kt=\br \Phi_1(0)|e^{{\rm i}(H^\dagger-H)\,t}
 |\Phi_2(0)\kt
  \neq
 \br \Phi_1(0)|\Phi_2(0)\kt\,.
 \label{neuni}
 \ee
Under certain subtle mathematical assumptions, fortunately, the
representation of the system may be changed in such a manner that
its evolution in time is made unitary again. In essence, one must
just abandon the traditional (i.e., the so called Dirac's
``transposition plus complex conjugation'') {\em special}
Hermitian-conjugation operation
 \be
 {\cal T}^{(Dirac)}:|\Phi(t)\kt\ \to \ \br \Phi(t)|
 \ee
and replace it by the {\em fully general},
arbitrary-metric-dependent version
 \be
 {\cal T}^{(\Theta)}:|\Phi(t)\kt\ \to \ \br \Phi(t)|\,\Theta\,.
 \ee
The details may be found in Ref.~\cite{SIGMA}.

The related replacement of the (by assumption, prohibitively
complicated) Eq.~(\ref{SEtaune}) by its (by assumption,
computationally friendly) alternative Eq.~(\ref{SEtautrane}) is
rendered consistent by the time-independence assumptions
$\mathfrak{h}\neq \mathfrak{h}(t)$ and $H\neq H(t)$. In what follows
we shall pay attention to the more general, time-dependent
crypto-Hermitian-Hamiltonian scenario which has been described in
paper I and in which $\mathfrak{h}= \mathfrak{h}(t)$, $H= H(t)$ and
$\Omega=\Omega(t)$.

\section{Manifestly time-dependent non-Hermitian quantum Hamiltonians
 \label{hoho} }

The method of simplification $\mathfrak{h} \to H$ of the
Hamiltonians as mediated by Eq.~(\ref{simplifya}) using non-unitary
$\Omega \neq 1/\Omega^\dagger$ cannot be transferred to the case of
manifestly time-dependent Hamiltonians. Still, the very idea itself
remains applicable. In a way described in our preceding paper I
\cite{timedep} one only has to rewrite Eq.~(\ref{simplifya})
accordingly,
 \be
 \mathfrak{h}(t) = \Omega(t)\,H(t)\,\Omega^{-1}(t)\,.
 \label{simplifybe}
 \ee
It is necessary to start from the time-dependent-Hamiltonian version
of the standard textbook Schr\"{o}dinger Eq.~(\ref{SEtaune}) without
any elementary solution,
 \be
 {\rm i}\p_t|\varphi(t)\kt = \mathfrak{h}(t)\,|\varphi(t)\kt\,,
 \ \ \ \ \ \
 |\varphi(t)\kt \in {\cal H}^{(P)}\,.
 \label{SEtaujo}
 \ee
Next,  we set
 \be
 |\varphi(t)\kt= \Omega(t)\,|\Phi(t)\kt\,,
 \ \ \ \ \ \
 \br \varphi(t)|= \br \Phi(t)|\Omega^\dagger(t)\,
 \ee
and, in the notation of Ref.~\cite{SIGMA}, define the {\em
auxiliary} ketkets and brabras,
 \be
 |\Phi(t)\kkt= \Omega^\dagger (t)\,|\varphi(t)\kt\,,
 \ \ \ \ \ \
 \bbr \Phi(t)|= \br \varphi(t)|\Omega(t)\,.
 \ee
This notation enables us to replace Schr\"{o}dinger
Eq.~(\ref{SEtaujo}) with hermitian $\mathfrak{h}(t)$ by the
following pair of its equivalent non-Hermitian descendants
 \be
 {\rm i}\p_t|\Phi(t)\kt = G(t)\,|\Phi(t)\kt\,,
 \ \ \ \ \ \
 |\Phi(t)\kt \in {\cal H}^{(F)}\,,
 \label{SEtaujoa}
 \ee
 \be
 {\rm i}\p_t|\Phi(t)\kkt = G^\dagger(t)\,|\Phi(t)\kkt\,,
 \ \ \ \ \ \
 |\Phi(t)\kkt \in {\cal H}^{(F)}\,
 \label{SEtaujobe}
 \ee
where we abbreviated $ G(t) = H(t) - \Sigma(t)$ with
 \be
 \Sigma(t)
 = {\rm i}\Omega^{-1}(t)\left [\partial_t\Omega(t)\right ]
 = {\rm i}\Omega^{-1}(t)\dot{\Omega}(t)\ \equiv \
 \Omega^{-1}(t)\,\sigma(t)\,\Omega(t)\,.
  \ee
A few further relevant remarks may be found in paper I.

\section{Simplification: constructive guarantee of the
time-independence of $ G(t)=G(0)$ \label{hohoho}}

The implementation costs of the generalization $\mathfrak{h} \to
\mathfrak{h}(t) \to G(t)\neq G^\dagger(t)$ as reviewed in preceding
section were most thoroughly discussed in Refs.~\cite{Bila}. The
author suggested that from a purely pragmatic perspective, our main
attention should be paid to the applications in which one is allowed
to work in an adiabatic approximation where the influence of
$\Sigma(t)$ may be neglected. One of such applications (viz., in
cosmology) has subsequently been outlined in Ref.~\cite{Bang}.

In our present text we do not intend to propose any approximations.
Rather, we shall follow the methodical guidance offered by
Ref.~\cite{Geyer}. In this setting one assumes, first of all, that
the operator $\mathfrak{h}(t)$ is, for virtually any purpose,
prohibitively complicated. This is accompanied by the second
assumption that there exists a non-unitary Dyson's map
$\Omega=\Omega(t)$ such that the solution of the mutually adjoint
Schr\"{o}dinger Eqs.~(\ref{SEtaujoa}) or (\ref{SEtaujobe}) becomes
{\em perceivably simpler} than the solution of their self-adjoint
predecessor Eq.~(\ref{SEtaujo}).

Next, we shall accept the most natural assumption that our choice of
$\Omega(t)$ is such that the new crypto-Hermitian generator $G(t)$
of time evolution becomes time-independent. Thus, we must show that
such an arrangement is possible and consistent and that it can lead
to the sufficiently persuasive simplification of the description of
the quantum system in question.

The latter requirement means that $G(t)=G(0)$ at all of the relevant
times. This would immediately imply the validity of the explicit and
compact exponential-operator formula for wave functions. Thus, for $
|\Phi(t)\kt  \in {\cal H}^{(F)}\,$ and $|\Phi(t)\kkt \in {\cal
H}^{(F)}\,$ we would have
 \be
  |\Phi(t)\kt=\exp (-{\rm i}G(0) \,t)\,|\Phi(0)\kt\,,
  \ \ \ \ \ \ \ \
  |\Phi(t)\kkt=\exp (-{\rm i}G^\dagger(0) \,t)\,|\Phi(0)\kkt\,.
 \label{taujoa}
 \ee
The manifestly guaranteed unitarity of the time evolution follows in
both the old (i.e., trivial-metric) and new (i.e., {\em ad
hoc}-metric) pictures. Indeed, having any product $\br
\varphi_1(t)|\varphi_2(t)\kt  \ \equiv \ \br \Phi_1(t) |\Theta(t)
|\Phi_2(t)\kt$ we may rewrite it, in the light of
Eq.~(\ref{taujoa}), in the equivalent form
 \be
  \bbr \Phi_1(t)|\Phi_2(t)\kt=\bbr \Phi_1(0)|\Phi_2(0)\kt=
 \br \varphi_1(0)|\varphi_2(0)\kt\,.
 \ee
Our task is reduced to the analysis of the existence of the
necessary time-dependent Dyson mapping $\Omega(t)$ such that it
satisfies our simplification requirements. In other words, we must
postulate the existence of the suitable time-dependent mapping
mediated by a not yet specified operator $\Omega(t)$ and by
Eq.~(\ref{simplifybe}) such that the Dyson-type time-dependent
transformation of the Hamiltonian operator $\mathfrak{h}(t) \to
H(t)$ {\em is} a simplification.

In the preparatory step it is  sufficient to guarantee (or assume)
such a simplification property at an initial instant $t=0$ {\em and}
at an infinitesimally shifted time $t=0+dt=\triangle>0$. This will
enable us to

\begin{itemize}

\item
select and fix one of many eligible \cite{Geyer} time-independent
operators $\Omega(0)$;

\item
evaluate the transformed, simplified operator
$H(0)=\Omega^{-1}(0)\mathfrak{h}(0)\Omega(0)$;

\item
select and fix one of the  operators $\Omega(\triangle)$;

\item
evaluate, with any predetermined precision, the time-independent
auxiliary operator $\dot{\Omega}(0)\approx
[\Omega(\triangle)-\Omega(0)]/\triangle +{\cal O}(\triangle^2)$;

\item
recall the appropriate definitions and specify operators $
\Sigma(0)$ and
 \be
 G(t)=G(0)=\Omega^{-1}(0)\mathfrak{h}(0)\Omega(0)
 -{\rm i}\Omega^{-1}(0)\dot{\Omega}(0)\,;
 \ee

\item
construct, ultimately, the time-evolving states in closed form
(\ref{taujoa}).

\end{itemize}

 \noindent
Our task is completed. Naturally, what is still missing here is a
constructive return to the original Hilbert space ${\cal H}^{(P)}$
which remains complicated. Whenever asked for, this step would
require the explicit reconstruction of the Dyson's operator
$\Omega(t)$ at all times. The necessary recipe will be outlined in
the next section.

\section{Discussion \label{hohohoho}}

The successful nuclear-physics tradeoff between the fermionic
Fock-space antisymmetrizations and the bosonic non-Hermiticity
complications has been described in Ref.~\cite{Geyer}. Similarly,
the manifest non-Hermiticity of certain toy-models in field theory
has been found a good price for the resulting feasibility of the
search for their discrete spectra \cite{Carl}.

These results should be perceived as a strong methodical support of
our present proposal of tradeoff between the loss of the elementary
time-evolution formula for the time-dependent Hermitian quantum
systems and the apparently non-unitary form of the simplified
crypto-Hermitian prescription (\ref{taujoa}). In order to make such
a tradeoff mathematically complete, we must return now to the
underlying postulate
 \be
 \partial_t\,G(t)=0
 \ee
which may be given the form $\dot{H}(t)=\dot{\Sigma}(t)$ or,
equivalently,
 \be
 {\rm i}\dot{\sigma}(t)={\rm i}\dot{\mathfrak{h}}(t)
 +\mathfrak{h}(t) \,\sigma(t)
 -\sigma(t)\,
 \mathfrak{h}(t)\,.
 \label{prvni}
 \ee
In the light of the above-mentioned definitions we may also write
down the second first-order differential equation
 \be
 {\rm i}\dot{\Omega}(t) = \sigma(t)\,\Omega(t)\,.
 \label{prvnice}
 \ee
At $t=0$ the latter relation specifies, first of all, the initial
value $\sigma(0)$ of the (not yet known) auxiliary operator function
$\sigma(t)$. This initial value just combines the above-specified
zero-time operators $\Omega(0)$ and $\dot{\Omega}(0)$. Subsequently,
the full reconstruction of the time-dependent operator $\sigma(t)$
must be performed via the linear differential Eq.~(\ref{prvni}).

In the final step, the resulting solution $\sigma(t)$ must be
inserted in Eq.~(\ref{prvnice}). The solution of the latter equation
will ultimately resolve the puzzle leading, at all the times $t$, to
the explicit form of the ``missing'' Dyson operator $\Omega(t)$.

\section{Summary \label{hohohohoho}}

In the majority of the existing practical applications of the
crypto-Hermitian representations of the operators of quantum
observables, the most difficult part of the constructions, viz., the
explicit determination of the Dyson mappings $\Omega$ is either
being declared redundant (and not performed at all) or found not too
essential (in such a case one only proceeds approximatively). Thus,
we are very rarely interested in the exact knowledge of operator
$\Omega$ or of its Hilbert-space-metric descendant
$\Theta=\Omega^\dagger\Omega$. For an illustration of this slightly
unexpected convention, it is sufficient to recollect that even for
one of the most popular crypto-Hermitian and ${\cal PT}-$symmetric
toy models using the imaginary cubic $H = -d^2/dx^2+{\rm i}\epsilon
x^3$, only the first three terms in the perturbation series for
$\Theta$ are known (cf. Ref.~\cite{cubic}). We may summarize that
generically, the crypto-Hermitian-representation approach to quantum
theory just works with an incomplete, reduced information about the
system in question. This feature of the method is one of its key
characteristics, concerning the majority of the applications of the
crypto-Hermitian quantum models, manifestly time-dependent or not.

There exist several ways towards the concrete implementations of
such an approach to quantum theory. For illustration let us just
recall the variational-method pattern used in the interacting boson
models of nuclear spectra (where one is {\em not} interested in the
construction of the wave functions {\em or} of any other observables
\cite{Geyer}), or the recipe applied to the most popular imaginary
cubic oscillator (where one selects just a very particular and, in
fact, unique mapping $\Omega$ which remains compatible with an
additional requirement of the observability of a charge ${\cal C}$
\cite{Carl}).

In our present text our considerations proceeded along the similar
lines. They were aimed at the maximal fructification and at an
explicit demonstration of the calculations-simplifying role of the
generic, {\em time-dependent} non-unitary Dyson's mappings
$\Omega(t)$. In a way complementing paper I we emphasized that in
the time-dependent cases such a mapping leads not only to the
replacement of a given phenomenological Hamiltonian
$\mathfrak{h}(t)$ by its ``instantaneous'', isospectral friendlier
partner $H(t)=\Omega^{-1}(t) \mathfrak{h}(t)\Omega(t)$ but also to
the possibility of the transfer of the evolution-generating role of
$\mathfrak{h}(t)$ to a pair of different and, incidentally,
particularly simple and, first of all, {\em manifestly
time-independent} operators $G$ and $G^\dagger$. These operators
were shown to appear in the respective partner Schr\"{o}dinger
equations (\ref{SEtaujoa}) and (\ref{SEtaujobe}). Both these
operators may be characterized by the hidden form of their
Hermiticity {\em as well as} by their time-independence. Leading to
the (perhaps, surprising?) closed exponential-operator form of the
evolution operators as well as to the hidden but, naturally,
necessary unitarity (or, if you wish, crypto-unitarity) of the
resulting quantum evolution law (\ref{taujoa}).

\subsection*{Acknowledgement}

Work supported by GA\v{C}R, grant Nr. P203/11/1433.

\end{document}